\documentclass[fleqn,twoside]{article}
\usepackage{amsmath}  
\usepackage{amssymb} 
\usepackage{espcrc2}

\usepackage{graphicx}
\usepackage[figuresright]{rotating}

\newcommand{\be}{\begin{equation}}
\newcommand{\ee}{\end{equation}}
\newcommand{\bes}{\begin{equation*}}
\newcommand{\ees}{\end{equation*}}
\newcommand{\slashme}[1]{\hspace{4pt}\slash \hspace{-6pt}#1}

\title{QCD at zero baryon density}

\author{Slavo Kratochvila\address[ETHZ]{Institute for Theoretical Physics, ETH Z\"{u}rich, CH-8093 Z\"{u}rich, Switzerland}\thanks{Talk presented by S. Kratochvila} and
        Philippe de
        Forcrand\addressmark[ETHZ]\address[CERN]{CERN, Theory Division, CH-1211 Geneva 23, Switzerland}}

\begin{document}

\markright{\hfill\rm CERN-TH/2003-222\quad\quad}

\begin{abstract}

While the grand canonical partition function $Z_{GC}(\mu)$ with chemical potential $\mu$
explicitly breaks the $Z_3$ symmetry with the Dirac determinant,
the canonical partition function at fixed baryon number $Z_C(B)$ is
manifestly $Z_3$-symmetric. We compare $Z_{GC}(\mu=0)$ and $Z_C(B=0)$
formally and by numerical simulations, in particular with respect
to properties of the deconfinement transition. Differences between the two ensembles, for physical observables
characterising the phase transition, vanish with increasing lattice size.
We show numerically that the free energy density is the same for both ensembles in the thermodynamic limit.

\vspace{1pc}
\end{abstract}

\maketitle

\section{INTRODUCTION}

\begin{figure}[htb] \label{fig:introduction_polloop}
\vspace{-0.55cm}

\begin{center}
\hspace*{-0.75cm} \includegraphics[width=9.0cm]{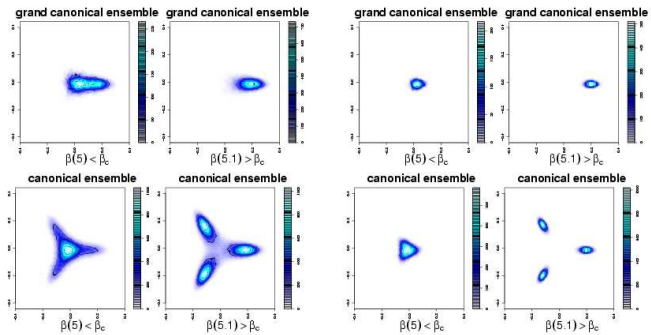}
\vspace{-1.35cm}

\caption{Distribution of the complex Polyakov loop in the grand
canonical (top) and canonical (bottom) ensembles. left: $4^3\times
4$, right: $6^3\times 4$. In the thermodynamic limit, the
distributions agree for both ensembles, up to two additional
$Z_3$-rotations in the canonical ensemble.}
\vspace{-1.20cm}

\end{center}
\end{figure}

The grand canonical partition function with respect to the quark number
\be \label{eq:introduction_ZGC}
Z_{GC}(\mu) = \int [DU] e^{-S_g[U]} \det M(U;\mu),
\ee
which is commonly used, explicitly breaks the $Z_3$-centre symmetry.
We compare it with the canonical partition function at fixed baryon number $B$
\be \label{eq:introduction_ZC}
Z_{C}(B) =  \frac{1}{2\pi}\int_{-\pi}^{\pi} \hspace{-0.25cm} d \bar \mu_I e^{-i 3 B \bar \mu_I} Z_{GC}(\mu = i \bar \mu_I T), \hspace{-0.05cm}
\ee
which preserves the $Z_3$-center symmetry. Here, we consider the case $\mu=B=0$ only.
While in one ensemble the expectation value of the Polyakov loop is non-zero at any temperature,
in the other it is always zero, as visible in Fig.~\ref{fig:introduction_polloop}.
Therefore, an important question arises:
Do both formulations agree in the thermodynamic limit for physically relevant
observables?
This is a subtle issue, leading to a controversy about
contributions of non-zero \emph{triality}\footnote[1]{ The difference
of the quark and anti-quark numbers modulo 3, $t \equiv (\#Q -
\#\bar Q) \mod 3$, is called the quark \emph{triality}. } states.
Refs.\cite{Oleszczuk:yg} suggest that non-zero triality
(``unphysical'', fractional $B$) states give a non-vanishing
contribution even in the thermodynamic limit,
and must be taken
into account. In contrast, \cite{Azcoiti:1998rj} shows, for QCD
with staggered fermions, that the only relevant sector of the
grand canonical $\mu=0$ description in the thermodynamic limit is
the $B=0$ sector (the free energy densities of both descriptions are
identical for $V \to \infty$).
In other words,  the canonical ensemble $Z_C(B=0)$, which contains only zero-triality states, describes the same
physics in the thermodynamic limit as the grand canonical ensemble $Z_{GC}(\mu=0)$,
which has contributions from all triality-sectors.
Therefore, non-zero triality states do not contribute in this limit.
By an explicit comparison of $Z_{GC}(\mu=0)$ with $Z_C(B=0)$, we hope to settle this discussion.

\section{THE CANONICAL ENSEMBLE}

\begin{figure}[tb]\label{fig:ensemble_phasediagram}
\begin{center}
    \includegraphics[width=5.0cm]{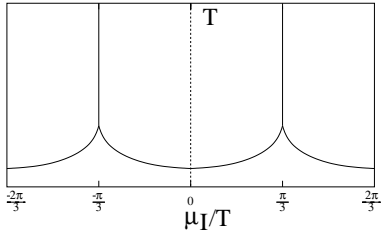}
\vspace{-0.75cm}

  \caption{Phase diagram of $Z_{GC}(i\mu_I)$ in the $(\mu_I,T)$-plane.
  The transitions (full lines) are first order for four flavours of staggered quarks with mass $m a=0.05$.}
\end{center}
\vspace{-0.75cm}

\end{figure}

We briefly recall the construction of the canonical partition function $Z_C(B)$.
First, one fixes the number of quarks $N= \int d^3 \vec x \; \bar{\psi}(\vec x) \; \gamma_0 \; \psi(\vec x)$ to $Q$
by inserting a $\delta$-function in the grand canonical
partition function:
\bes
Z_{C}(Q) = \int [DU][d\bar{\psi}][d\psi] e^{-S_g[U]-S_F[U,\bar{\psi},\psi]} \delta\left(N-Q\right)\;.
\ees
The $\delta $-function admits a Fourier representation:
\begin{gather*}
Z_{C}(Q) = \frac{1}{2\pi}\int_{-\pi}^\pi d\bar \mu_I  e^{-i Q \bar \mu_I}  \times  \\
\times \int [DU][d\bar{\psi}][d\psi] e^{-S_g[U]-S_F[U,\bar{\psi},\psi] + i  \bar \mu_I N }
\end{gather*}
with the new variable $\bar \mu_I$.
One recognizes $\mu_I = \bar \mu_I T$  as an imaginary chemical potential, so that
\bes
Z_{C}(Q) = \frac{1}{2\pi} \int_{-\pi}^\pi d\bar \mu_I  e^{-i Q \frac{\mu_I}{T} } Z_{GC}(i\mu_I ).
\ees
The grand canonical partition function $Z_{GC}(i\mu_I)$ as a function of an imaginary chemical potential
has interesting properties \cite{Roberge:mm}, see Fig.~\ref{fig:ensemble_phasediagram}.
It is even and $2\pi T/3$-periodic in $\mu_I$.
For our purposes, the most important consequence is, that $Z_{C}(Q  \neq 0 \mod 3) = 0$.
Therefore, we define the baryon number $B \equiv 3 Q$ and end up with Eq.(\ref{eq:introduction_ZC}).

\section{RESULTS}

We simulate QCD with $N_f=4$ flavours of staggered quarks of mass $m a=0.05$, using Hybrid
Monte Carlo.
We measure the Polyakov loop, the plaquette and the chiral condensate
on lattices with extents $4^3$, $6^3$ and $8^3
\times 4$ at seven different $\beta$'s, from $\beta_1=4.9$ to
$\beta_7=5.1$. We analyse the results using multihistogram reweighting \cite{Ferrenberg:yz}.

In order to sample the $B=0$ canonical ensemble, we supplement the
ordinary Hybrid Monte Carlo at fixed $\mu_I$ with a Metropolis update $\mu_I \rightarrow \mu_I'$, with
acceptance $\min\left(1,\frac{\det^{N_f}(  \slashme{D} (\mu_I') + m)
}{\det^{N_f}(  \slashme{D} (\mu_I) + m)}\right)$.
The determinant ratio is evaluated with a stochastic estimator,
namely $\langle e^{- |( \slashme{D} (\mu_I') + m)^{-N_f/2} ( \slashme{D}
(\mu_I) + m)^{N_f/2} \eta| ^2 + |\eta|^2 }   \rangle_\eta$ where
$\eta$ is a gaussian complex vector. In addition,
one can perform a ``$Z_3$ move'' at any time (with acceptance $1$):
${\mu_I \to \mu_I \pm \frac{2\pi T}{3}}$, ${U_4(\vec x,t=0) \to U_4(\vec x,t=0)e^{\mp i
\frac{2\pi}{3}},\;\forall \vec x}$.
\begin{figure}[htb] \label{results_chiral}
\vspace{-0.75cm}

\begin{center}
\hspace{-0.75cm} \includegraphics[width=2.925cm,angle=-90]{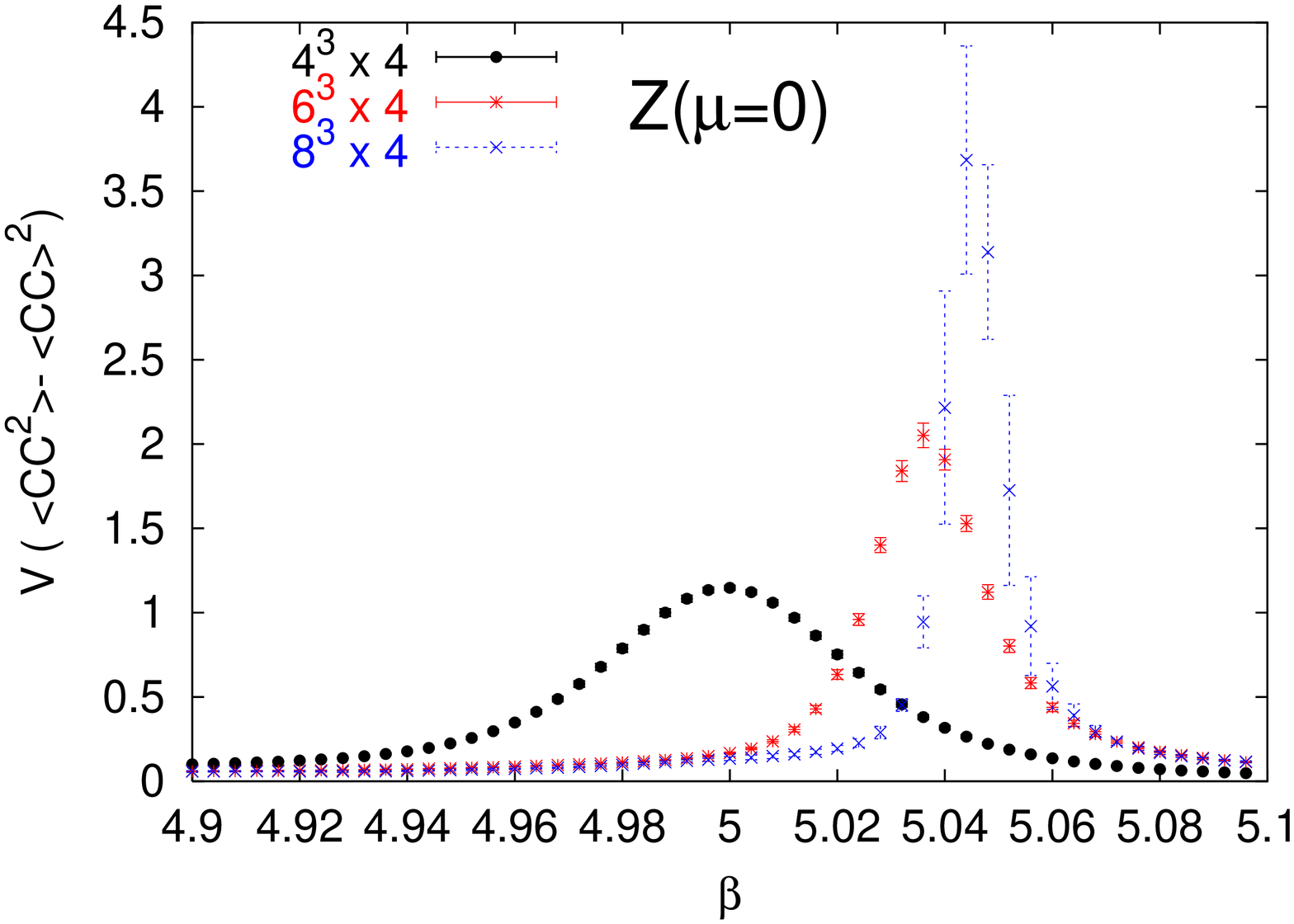}
\hspace{-0.35cm} \includegraphics[width=2.925cm,angle=-90]{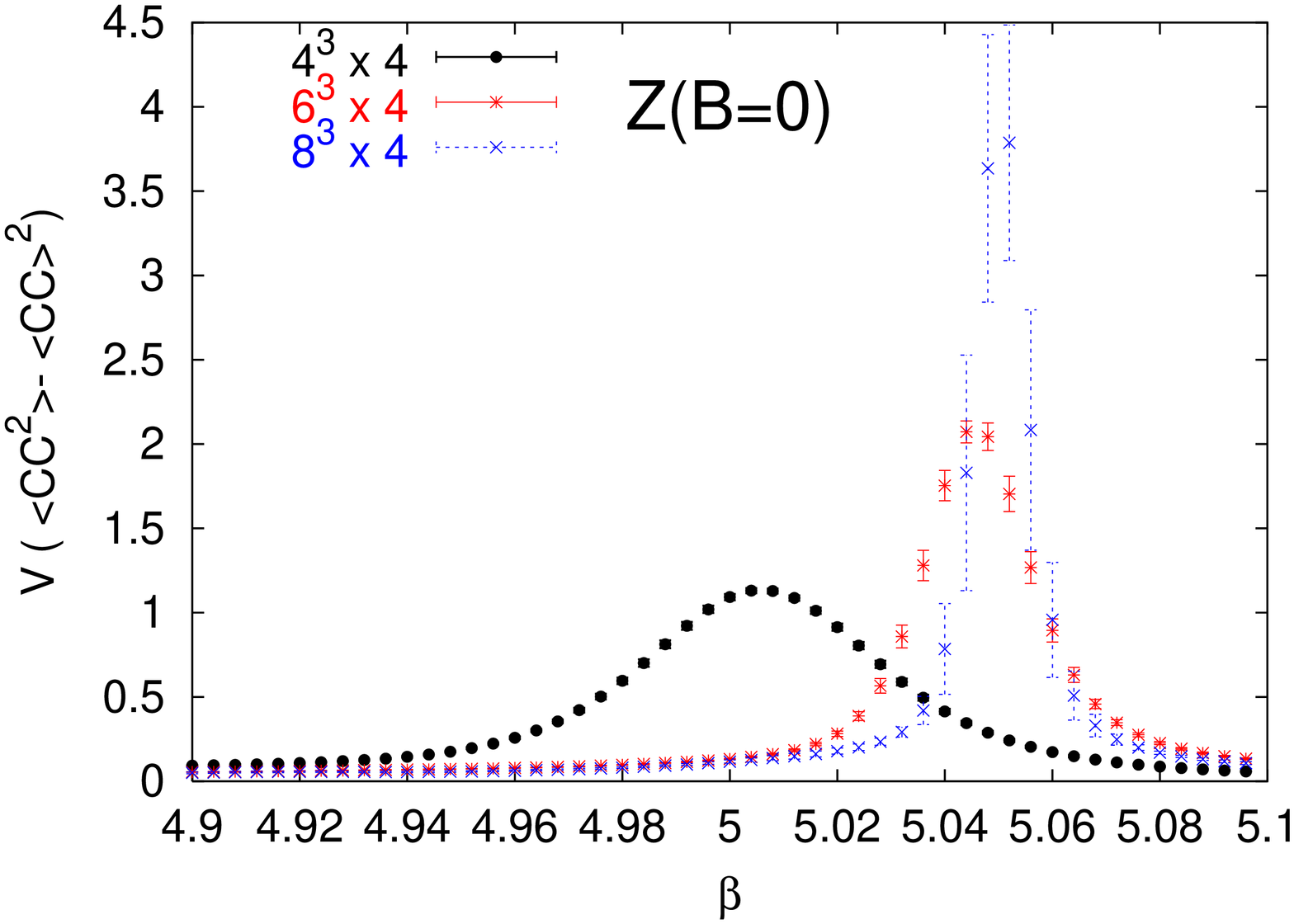}\hspace{-0.75cm}
\vspace{-0.75cm}

\caption{Susceptibility of $\bar{\psi} \psi$ on all analysed lattices and both
ensembles. left: $Z_{GC}(\mu=0)$, right: $Z_C(B=0)$. Even for the smallest, $4^4$, lattice,
differences are barely visible.}
\end{center}
\vspace{-0.75cm}

\end{figure}

Fig.~3 shows the susceptibility of the chiral condensate $\bar{\psi} \psi$ in both ensembles.
The peak shifts slightly in $\beta$ for the $4^3 \times 4$ lattices. For larger lattice sizes, the difference disappears
and both ensembles indicate the same critical $\beta_c$ already for the $8^3 \times 4$-lattices. We observe the same behaviour for the average plaquette.
This is our first evidence to answer the question about the influence of non-zero triality states: In the thermodynamic limit,
non-zero triality sectors are suppressed and do not affect the results. The effect of non-zero triality sectors
is smaller than the statistical errors.
\begin{figure}[htb]\label{results_binder}
\vspace{-0.25cm}

\begin{center}
\hspace{-0.75cm} \includegraphics[width=2.925cm,angle=-90]{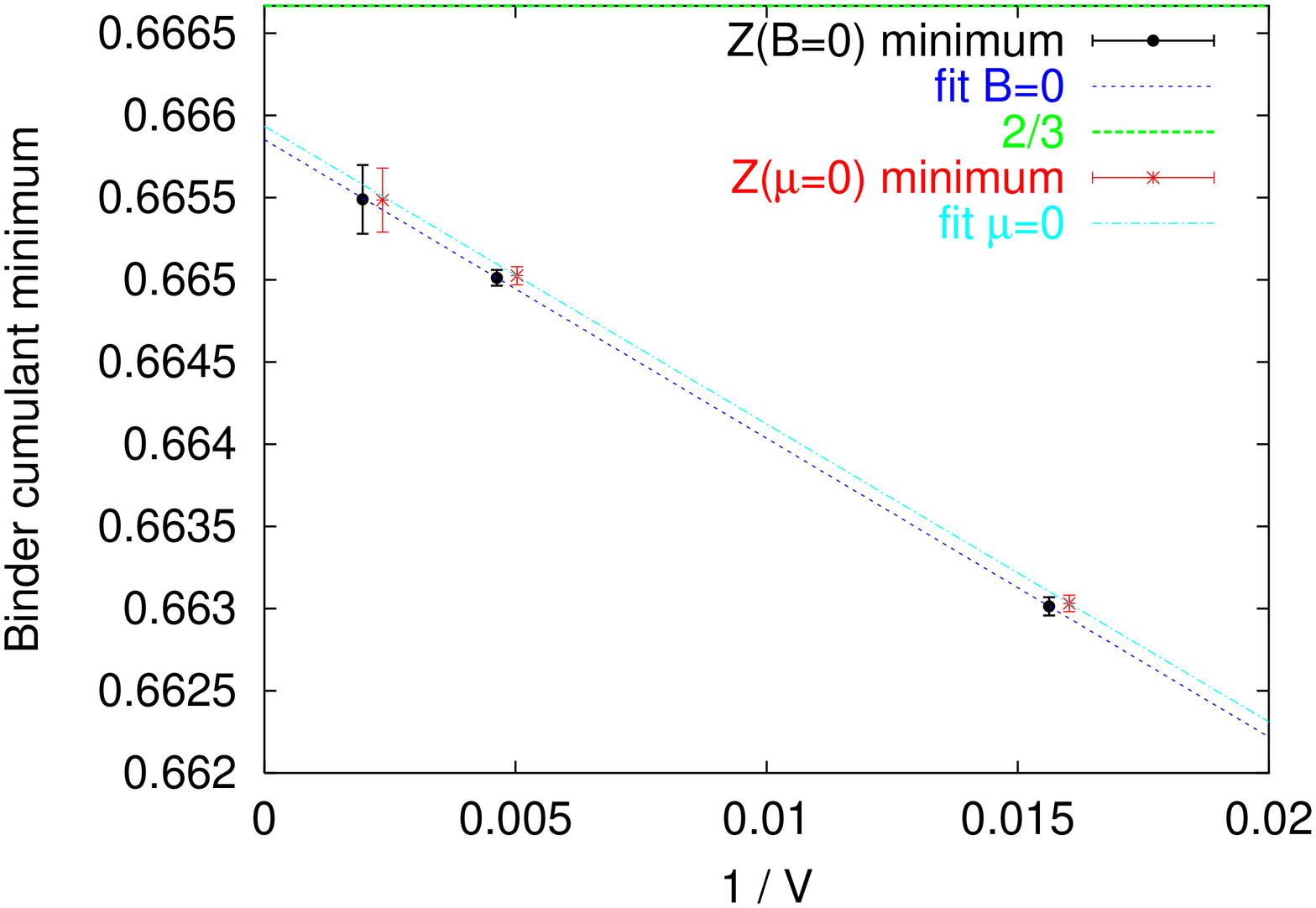}
\hspace{-0.35cm} \includegraphics[width=2.925cm,angle=-90]{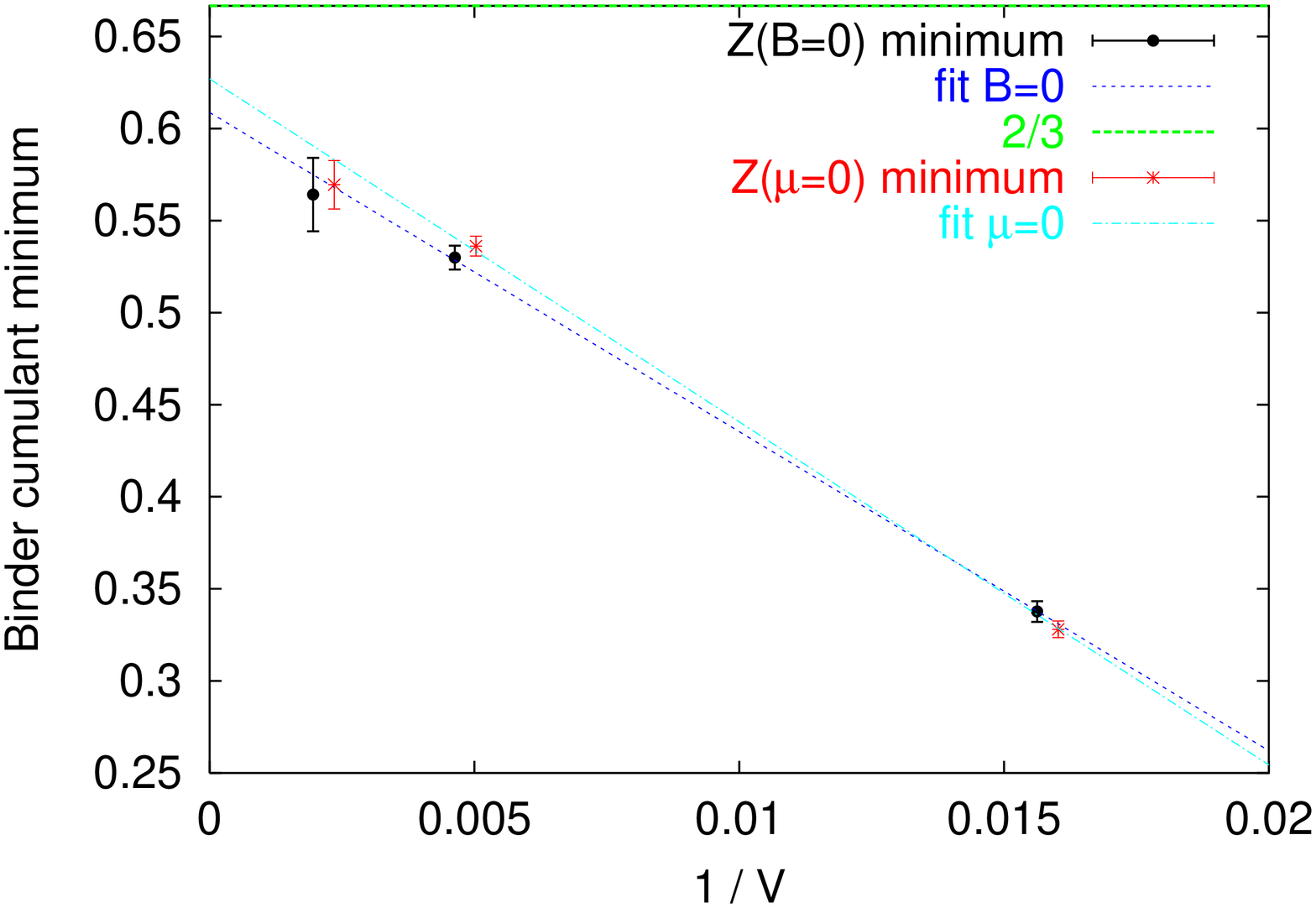}\hspace{-0.75cm}
\vspace{-0.75cm}

\caption{Binder cumulant minimum
versus inverse volume for
both ensembles (slightly shifted). left: average plaquette, right:
chiral condensate. The thermodynamic extrapolation does not tend to
$\frac{2}{3}$, indicating a first order transition.}
\end{center}
\vspace{-0.75cm}

\end{figure}

In Fig.~4 we show a finite size analysis of observables in the two ensembles.
We plot the minimum of the Binder cumulant
$C_B({\cal O}) = 1 - \frac{1}{3} \frac{ \langle  {\cal O}^4  \rangle}{\langle {\cal O}^2 \rangle^2}$
as a function of the inverse volume $1/V$. For both the average plaquette and the chiral condensate, the thermodynamic extrapolation
does not tend to $\frac{2}{3}$ (top-limit of the plot) - indicative of a first order phase transition
(as known from the literature \cite{Fukugita:1986rr}). Furthermore, the slopes of the linear fit are identical in both ensembles within statistical errors.
Therefore, we cannot even claim smaller finite size effects in one of the two ensembles.
\begin{figure}[htb]\label{results_chempot}
\vspace{-0.75cm}

\begin{center}
\hspace{-0.75cm} \includegraphics[width=2.925cm,angle=-90]{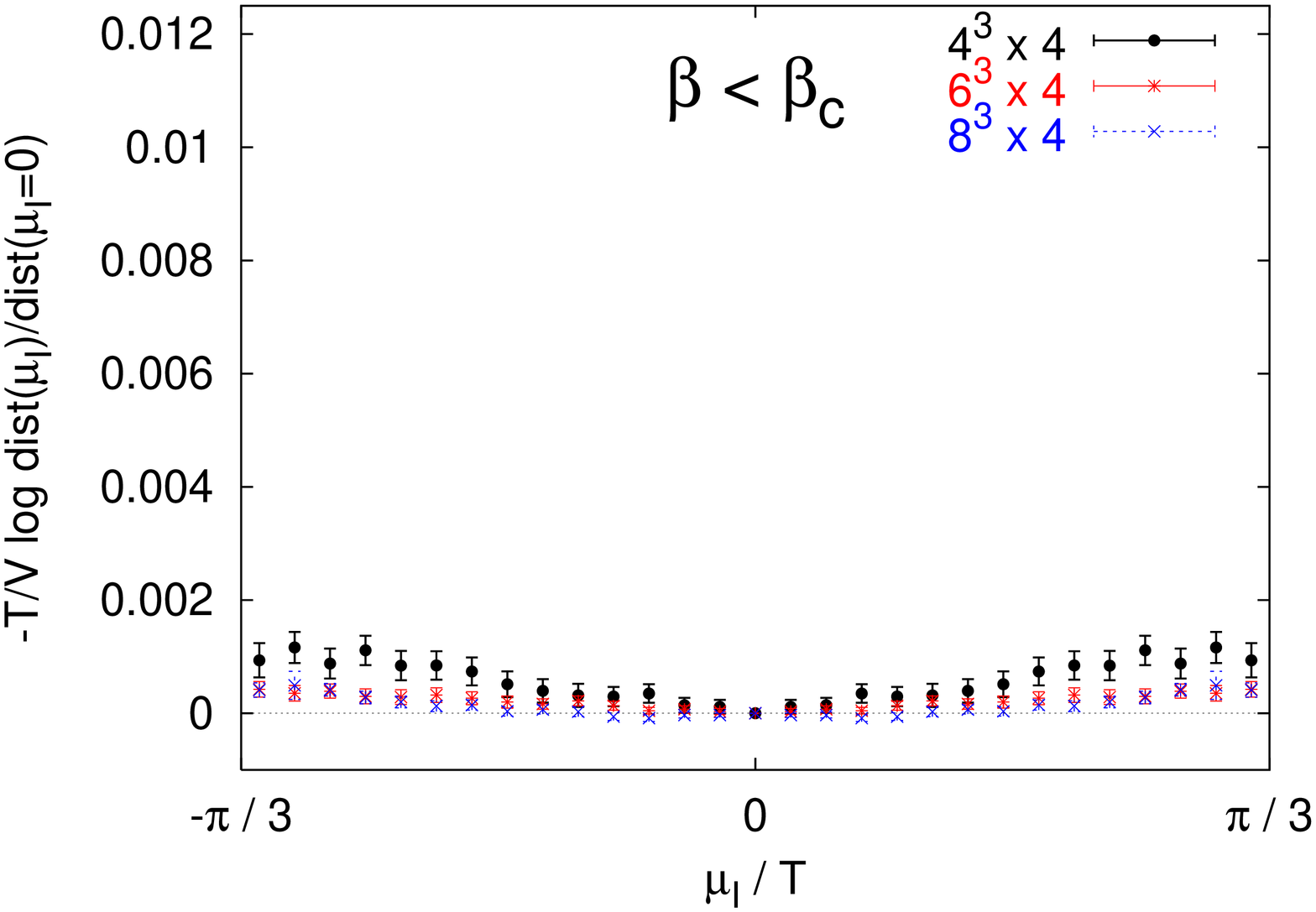}
\hspace{-0.35cm} \includegraphics[width=2.925cm,angle=-90]{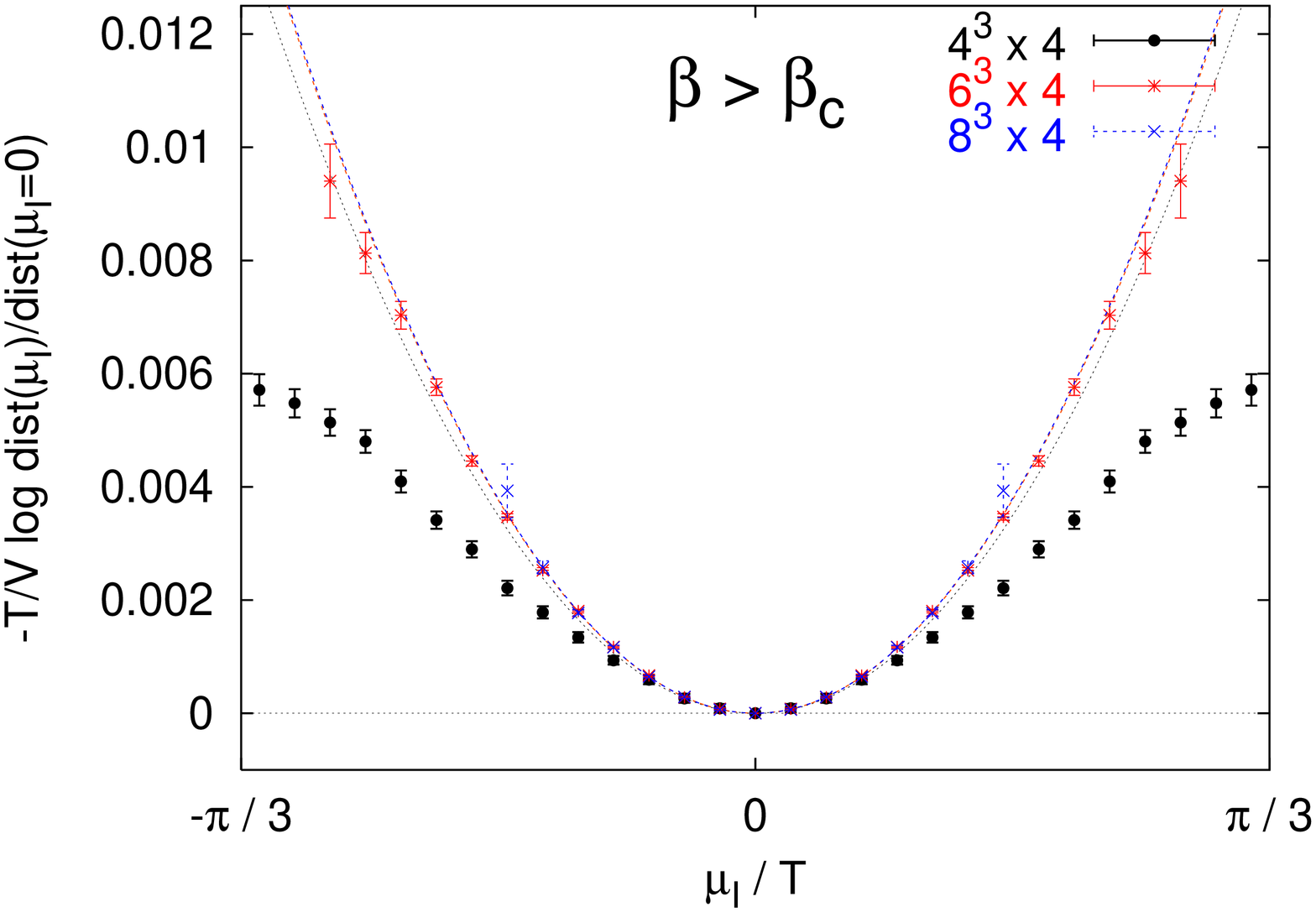}\hspace{-0.75cm}
\vspace{-0.75cm}

\caption{Free energy density as a function of $\mu_I$. left side: $\beta=4.95 < \beta_c$, right: $\beta=5.10> \beta_c$. In the latter case,
a parabola nicely describes the data for the larger lattices.}
\end{center}
\vspace{-0.75cm}

\end{figure}

In Fig.~5 we show the free energy density
versus $\mu_I$ for both $\beta < \beta_c$ and $\beta > \beta_c$.
In both cases, we clearly observe a minimum at $\mu_I=0$.
Therefore, in the thermodynamic limit, only $\mu_I=0 \mod \frac{ 2
\pi T}{3}$ will survive. This establishes the equivalence of $Z_C(B=0)$ with
$Z_{GC}(\mu=0)$. For $\beta=4.95 < \beta_c$, we observe that the
free energy density at $\mu_I = \pm \pi T/3$ seems to have zero
derivative, indicating a crossover (as expected from the phase
diagram Fig.~\ref{fig:ensemble_phasediagram}). Instead, for
$\beta=5.10> \beta_c$ we expect a cusp to develop, due to the
first order phase transition. Indeed, it appears likely as the
volume increases. A simple parabola gives a good description of
the data, with a curvature in line with expectations from high
temperature \cite{Karsch:2003zq}.

\section{CONCLUSION}

The thermodynamic limits of $Z_{GC}(\mu=0)$ and $Z_C(B=0)$ seem to agree with respect to the deconfinement phase transition.
The effect of non-zero triality sectors is smaller than the statistical errors, already for $8^3 \times 4$ lattices.
Therefore, these states can be included or excluded without affecting the results in the thermodynamic limit.
Moreover, finite size effects are equivalent even on the smallest, $4^3 \times 4$, lattice.

The canonical formulation requires a centre-symmetric formulation of QCD, which can be achieved very simply with negligible
computer overhead, by adding a single d.o.f.~$\mu_I$ updated by Metropolis.
Hence, the presence of a fermionic determinant does not necessarily break the centre-symmetry.

Note, that another grand canonical partition function $Z_{GC}(\mu)$ can be built from the $Z_C(B)$'s,
which is completely centre-symmetric. This partition function will give
identical expectation values to the usual $Z_{GC}(\mu)$, except for centre-sensitive observables (Polyakov loop),
whose different behaviour can be understood by invoking spontaneous center-symmetry breaking \cite{SKPdF}.

\markright{}

\end{document}